\documentclass[12pt,a4paper]{article}
\usepackage[cp866]{inputenc}
\usepackage[dvips]{epsfig}
\usepackage[english]{babel}
\usepackage[]{caption}

\begin{document}
\author{\bf Yu.A. Markov$\!\,$\thanks{e-mail:markov@icc.ru}
$\,$and M.A. Markova$^*$}
\title{Scattering of high-energy partons off ultrasoft\\
gluon fluctuations in hot QCD plasma\\
and energy losses}
\date{\it $^{\ast}\!$Institute for System Dynamics and Control Theory\\
Siberian Branch of Academy of Sciences of Russia,\\
P.O. Box 1233, 664033 Irkutsk, Russia}
\thispagestyle{empty}
\maketitle{}

\def\theequation{\arabic{section}.\arabic{equation}}

\[
{\bf Abstract}
\]
Within the framework of the effective theory for ultrasoft field modes a
problem of interacting an energetic color particle with ultrasoft fluctuations
in hot gluon plasma is considered. The procedure of calculation of certain
effective current generating the process of interaction is proposed, and the
problem of its gauge independence is discussed. The application of
the theory developed to the problem of energy losses of the energetic
color particle propagating
through hot QCD-medium is given. It is shown that the perturbation approach
for calculation of energy losses breaks down at the ultrasoft momentum
scale of plasma excitations.

{\sl PACS:} 12.38.Mh, 24.85.+p, 11.15.Kc

\newpage

\section{\bf Introduction}
\setcounter{equation}{0}

In this work we study the problem of scattering of energetic
color particles (quarks, gluons or more generally -- partons) off plasma
fluctuations at the ultrasoft momentum scale
($p_0 \sim g^4T,\;|{\bf p}|\sim g^2T$, where $T$ is temperature of the
system and $g$ is the coupling constant) in hot gluon plasma.
The correct effective theory at the ultrasoft
momentum scale is generated by the linearized Boltzmann-Langevin equation
in the form first proposed by B\"odeker \cite{bodeker1}.
Then this equation was obtained
by a number of authors \cite{arnold1, valle, litim1, blaizot1} within the
framework of different approaches. This transport equation includes
the collision term for color relaxation and the Gaussian noise term that
keeps the ultrasoft modes in thermal equilibrium.

The way for solving the problem stated is based on general ideas
suggested in our previous work \cite{markov1} dealing with the
similar problem of scattering of high-energy particles off the
plasma fluctuations at the soft momentum scale ($p_0\sim
gT,\;|{\bf p}|\sim gT$). Following the idea of the work
\cite{markov1} we supplement the Boltzmann-Langevin equation by
the Wong equation \cite{wong} describing the precession of the
classical color charge $Q=(Q^a),\,a=1,\ldots, N_c^2-1$ of the
energetic parton propagating through QCD-medium. The scattering
process is generated by certain effective current. The calculation
of this current is crucial step of the formalism under
consideration. However in contrast to our previous work
\cite{markov1} the effective current here represents the expansion
not in powers of the free gauge field $A_{\mu}^{(0)}(X)$ (that at
this momentum scale is strongly damping by virtue of collisions),
but in powers of the noise term $\nu(X,{\bf v})$ and also initial
value of the color charge $Q_0$ of the energetic parton.
We work to lowest order in the color charge $Q_0$ of the parton,
but to all orders in the noise term, since the external parton is
a truly perturbative object, whereas the thermal color fluctuations
at the ultrasoft scale $g^2T$ are non-perturbative \cite{linde}.
We apply the current
approach to study of the energy losses of the energetic parton
induced by scattering off the ultrasoft gluon fluctuations. We
show that for weak gauge coupling this contribution to the total
balance of the energy losses can be neglected, although here,
there is some vagueness of the exact estimation of this type of
energy loss.

This paper is organized as follows. In Section 2, the
convention and notation used in this paper are summarized. In
Section 3, the basic nonlinear integral equation for gauge potential
is written out and the algorithm of the successive calculation of
certain effective current generating the interaction process
of the ultrasoft plasma fluctuations with the energetic color particle is
proposed. In Section 4, we discuss the problem of gauge independence
of matrix element for the simplest scattering process. In
Section 5, the energy losses caused by the scattering off ultrasoft
gluon fluctuations in lower orders in powers of the noise term is
analyzed.

\section{Initial equations}
\setcounter{equation}{0}

We use the metric $g^{\mu \nu} = diag(1,-1,-1,-1)$, choose units such
that $c=k_{B}=1$ and note $X=(X_0,{\bf X}), \,p=(p_0\equiv\omega,{\bf p})$ etc.

At the space-time scale $X\gg (gT)^{-1}$ the ultrasoft fluctuations of
the gluon color density\footnote{Here, we employ the parametrization
\cite{blaizot1} for deviation of the gluon density from the equilibrium
$\delta N({\bf k},X) = -g\,W(X,{\bf v})\,dN(|{\bf k}|)/d|{\bf k}|$,
where $N(|{\bf k}|)$ is the Bose-Einstein distribution.
More general expression for
parametrization of off-equilibrium fluctuations is considered in
Ref.\,\cite{markova}.} $W(X,{\bf v})=W^a(X, {\bf v})T^a$
$((T^a)^{bc}\equiv -if^{abc})$ satisfies the Boltzmann-Langevin
equation \cite{bodeker1}
\begin{equation}
[v \cdot D_X, W(X,{\bf v})] = -\, {\bf v}\cdot {\bf E}(X)
-{\rm C}[W](X,{\bf v}) + \nu(X,{\bf v}).
\label{eq:2q}
\end{equation}
Here, $D_{\mu} = \partial_{\mu} + igA_{\mu}(X)$ is the covariant derivative;
$[\,,\,]$ denotes the commutator; ${\bf E} (X) = {\bf E}^a(X) T^a$ is the
chromoelectric field and
${\rm C}[W]$ is linearized collision term\footnote{Hereafter the square
brackets denote functional dependence.} defined by the expression
\begin{equation}
{\rm C}[W](X,{\bf v}) \equiv \hat{\rm C}W(X,{\bf v}) =
\int\!\frac{d\Omega_{{\bf v}^{\prime}}}{4\pi}\,{\cal C}({\bf v},{\bf v}^{\prime})
W(X,{\bf v}^{\prime})
\label{eq:2w}
\end{equation}
with the collision kernel
\begin{equation}
{\cal C}({\bf v},{\bf v}^{\prime})=
\gamma \delta^{(S^2)}\!({\bf v} - {\bf v}^{\prime})-m_D^2\,
\frac{g^2N_cT}{2}\,\Phi ({\bf v}\cdot{\bf v}^{\prime}).
\label{eq:2e}
\end{equation}
In the last expression $\delta^{(S^2)}\!({\bf v}-{\bf v}^{\prime})$ is the
delta-function on the unit sphere,
\[
\Phi ({\bf v}\cdot{\bf v}^{\prime}) \simeq
\frac{2}{\pi^2m_D^2}\,
\frac{({\bf v}\cdot{\bf v}^{\prime})^2}
{\sqrt{1- ({\bf v}\cdot{\bf v}^{\prime})^2}}\,\ln\left(\frac{1}{g}\right),
\quad m_D^2= \frac{1}{3}\,g^2N_cT^2
\]
within logarithmic accuracy and
\[
\gamma = m_D^2\,\frac{g^2N_cT}{2}\!
\int\!\frac{d\Omega_{{\bf v}^{\prime}}}{4\pi}\,
\Phi ({\bf v}\cdot{\bf v}^{\prime})
\]
is the damping rate for hard transverse gluon with velocity ${\bf v}$.
Furthermore the function $\nu(X,{\bf v})=\nu^a(X,{\bf v})T^a$ is the noise term
with the noise-noise correlation function
\begin{equation}
\ll\! \nu^a(X,{\bf v}) \nu^b(X^{\prime},{\bf v}^{\prime})\!\gg\,\,=
\frac{2T}{m_D^2}\,{\cal C}({\bf v},{\bf v}^{\prime})
\delta^{ab}\delta^{(4)}\!(X-X^{\prime}).
\label{eq:2r}
\end{equation}
The color current in the terms of the function $W(X,{\bf v})$ is
\begin{equation}
j_{\mu}(X)=m_D^2\!\int\frac{d\Omega_{\bf v}}{4\pi}\,
v_{\mu} W(X,{\bf v}).
\label{eq:2t}
\end{equation}

Following Blaizot and Iancu \cite{blaizot2} we present the function
$W(X,{\bf v})$ as the sum of two parts
\[
W(X,{\bf v}) = W^{\,{\rm ind}}(X,{\bf v}) + {\cal W}(X,{\bf v}),
\]
where $W^{\rm ind}(X,{\bf v})$ is the solution of Eq.\,(\ref{eq:2q}) in the
absence of the noise $\nu(X,{\bf v})$
\begin{equation}
[v \cdot D_X, W^{\,{\rm ind}}(X,{\bf v})] =
-\,{\bf v}\cdot {\bf E}(X) - \hat{\rm C}W^{\,{\rm ind}}(X,{\bf v}),
\label{eq:2y}
\end{equation}
and ${\cal W}(X,{\bf v})$ is a fluctuating piece satisfying
\begin{equation}
[v \cdot D_X,{\cal W}(X,{\bf v})] =
-\,\hat{\rm C}{\cal W}(X,{\bf v})+\nu(X,{\bf v}).
\label{eq:2u}
\end{equation}
Thus, ${\cal W}(X,{\bf v})$ is proportional to $\nu$ and in general also
depends on the gauge field $A_{\mu}(X)$. The color current (\ref{eq:2t}) is
decomposed as
\begin{equation}
j_{\mu}(X)=m_D^2\!\int\frac{d\Omega_{\bf v}}{4\pi}\,
v_{\mu}\Bigl(W^{\,{\rm ind}}(X,{\bf v}) + {\cal W}(X,{\bf v})\Bigr)\equiv
j_{\mu}^{\;{\rm ind}}(X) + \zeta_{\mu}(X),
\label{eq:2i}
\end{equation}
where $\zeta_{\mu}(X)$ is the fluctuation current acting as the noise
term in the Yang-Mills equation
\begin{equation}
[D^{\nu},F_{\mu \nu}(X)] +
\xi^{-1} \partial_\mu \partial^\nu A_{\nu}(X) =
j_{\mu}^{\;{\rm ind}}(X) + \zeta_{\mu}(X).
\label{eq:2o}
\end{equation}
Here, $F_{\mu \nu}=F_{\mu \nu}^{a}t^{a}$ is the field strength tensor
and $\xi$ is the gauge parameter fixing a
covariant gauge.

If there is a moving energetic color charged parton (quark or gluon)
in the hot gluon plasma, then on the right-hand side of field
equation (\ref{eq:2o}) it is necessary to
add appropriate color current $j_{Q\mu}(X)$. One expects the world line of
this parton to obey the classical trajectory in the manner of Wong \cite{wong}.
In the leading order in the coupling constant the
color current of the energetic color parton has the form \cite{markov1}
\begin{equation}
j_Q^{a\mu}=g\check{v}^{\mu}U^{ab}(t,t_0)Q_0^b\,{\delta}^{(3)}
({\bf x}-\check{\bf v}t),\quad
\check{v}^{\mu}=(1,\check{\bf v}),
\label{eq:2p}
\end{equation}
where
\[
U(t,t_0) = {\rm T}\exp\{-ig\!\int_{t_0}^t\!
(\check{v}\cdot A^a(\tau,\check{{\bf v}}\tau))T^ad\tau\}
\]
is the evolution operator taking into account the color precession along the
parton trajectory, $\check{\bf v}$ is the velocity of the energetic parton and
$Q_0^a$ is initial value of its color charge.

\section{\bf Construction of effective current}
\setcounter{equation}{0}

Now we are in position to construct the effective theory of nonlinear
interaction of the energetic color particle with ultrasoft gluon
fluctuations in spirit of the work \cite{markov1}. As was mentioned in
Introduction, the main point here,
is deriving certain effective current generating this process
of nonlinear interaction. For this purpose, at first we represent the
solutions  $W^{\,{\rm ind}}(x,{\bf v})$ and ${\cal W}(x,{\bf v})$ of
dynamical equations (\ref{eq:2y}), (\ref{eq:2u}) as a formal expansion
in powers of interaction field $A_{\mu}(X)$. Rewriting the Yang-Mills equation
(\ref{eq:2o}) in the momentum
space\footnote{We add the current of the hard color charged parton
(\ref{eq:2p}) to the right-hand side of Eq.\,(\ref{eq:2o}).} and taking into
account above-mentioned expansions, we
obtain the basic nonlinear integral equation for the gauge potential
$A_{\mu}(p)$ at the ultrasoft momentum scale
$(p_0\sim g^4T, |{\bf p}|\sim g^2T)$
\begin{equation}
\,^{\ast}{\cal D}^{-1}_{\mu \nu}(p)A^{a\nu}(p)=
-j_{\mu}^{\,({\rm tot})a}[A,\nu](p),
\label{eq:3q}
\end{equation}
where an expression for total current in representation of the interaction
field reads
\begin{equation}
j_{\mu}^{\,({\rm tot})a}[\nu,A](p)=
j_{\mu}^{{\rm ind}\,a}[A](p) + j_{Q\mu}^{a}[A](p) +
\zeta_{\mu}^{a}[A,\nu](p).
\label{eq:3w}
\end{equation}
Note that in deriving Eq.\,(\ref{eq:3q}) the part linear in
$A_{\mu}(p)$ of the induced current (that is, the one involving the
polarization tensor $\Pi_{\mu \nu} (p)$) has been extracted out
from $j_{\mu}^{\,({\rm tot})a}$ and included in the effective
inverse propagator on the left-hand side of (\ref{eq:3q}).
Conversely, the tree-level 3-gluon and 4-gluon vertices which were
originally presented on the left-hand side of Eq.\,(\ref{eq:2o})
have been now reabsorbed into the induced current on the
right-hand side of Eq.\,(\ref{eq:3q}). Inverse gluon propagator on
the left-hand side of Eq.\,(\ref{eq:3q}) is defined by expression
\begin{equation}
\,^{\ast}{\cal D}_{\mu \nu}(p)= -
P_{\mu \nu}(p) \,^{\ast}\!\Delta^t(p) -
Q_{\mu \nu}(p) \,^{\ast}\!\Delta^l(p) +
\xi D_{\mu \nu}(p)\Delta^0(p),
\label{eq:3e}
\end{equation}
where $\Delta^0(p)=1/p^2$; $\!\,^{\ast}\!\Delta^{t,\,l}(p) =
1/(p^2 - \Pi^{t,\,l}(p)), \,
\Pi^t(p) = \frac{1}{2} \Pi^{\mu \nu}(p) P_{\mu \nu}(p),$ and
$\Pi^l(p) = \Pi^{\mu\nu}(p)Q_{\mu\nu}(p)$ with ultrasoft gluon self-energy
$\Pi_{\mu\nu}(p)$ \cite{blaizot3}
\[
\Pi_{\mu\nu}(p)=m_D^2\left\{
-g_{\mu 0}g_{\nu 0} + \omega\!
\int\frac{d\Omega_{\bf v}}{4\pi}
\int\frac{d\Omega_{{\bf v}^{\prime}}}{4\pi}\,
\Bigl\langle v\Big|\,v_{\mu}\frac{1}{v\cdot p +i\hat{\rm C}}
\,v_{\nu}\,\Big|v^{\prime}\Bigr\rangle
\right\}.
\]
The explicit form of the Lorentz matrices $P_{\mu\nu},\,Q_{\mu\nu}$ and
$D_{\mu\nu}$ is given in Appendix. The symbols $\langle v \vert$ and
$\vert v^{\prime} \rangle$ in the 
definition of $\Pi_{\mu \nu} (p)$
possess the following properties
\[
\Bigl\langle v \Bigm| v^{\prime} \Bigr\rangle=
\delta^{(S^2)}\!({\bf v} - {\bf v}^{\prime}),
\quad
\Bigl\langle v \Bigm|\hat{\rm C}\Bigm|v^{\prime}\Bigr\rangle =
\Bigl\langle v^{\prime} \Bigm|\hat{\rm C}\Bigm|v\Bigr\rangle =
{\cal C} ({\bf v},{\bf v}^{\prime}),
\quad
v_{\mu} \Bigl|v^{\prime}\Bigr\rangle =
v_{\mu}^{\prime} \Bigl|v^{\prime}\Bigr\rangle
\]
and satisfy the ``completeness relation''
\[
\int\frac{d\Omega_{\bf v}}{4\pi}\,
\Bigl|v\Bigr\rangle\, \Bigl\langle v \Bigr| = 1.
\]
In the works \cite{arnold2, guerin1} explicit analytical expressions
for functions $\Pi^{t}(p)$ and $\Pi^{l}(p)$ in the terms of the continued
fractions are given.

The induced color current on the right-hand side of Eq.\,(\ref{eq:3w}) by
virtue of the Boltzmann equation (\ref{eq:2y}) is expressed as
\begin{equation}
j_{\mu}^{{\rm ind}\,a}[A](p)=\sum_{s=2}^{\infty} j_{\mu}^{{\rm ind}(s)a}
(A,\ldots,A),
\label{eq:3r}
\end{equation}
where
\[
j_{\mu}^{{\rm ind}(s)a}(A,\ldots,A)
=  \frac{1}{s!}\,g^{s-1}\!\!\int\!\!
\,^{\ast}\Gamma^{a a_1\ldots a_s}_{\mu\mu_1\ldots\mu_s}(p,-p_{1},\ldots,-p_{s})
A^{a_1\mu_1}(p_{1})A^{a_2\mu_2}(p_{2})\ldots A^{a_s\mu_s}(p_{s})
\]
\begin{equation}
\times\,
{\delta}^{(4)}\!\biggl(p - \sum_{i=1}^{s}p_{i}\biggr)\prod_{i=1}^{s}dp_{i}.
\label{eq:3t}
\end{equation}
In the last expression the coefficient functions
$\,^{\ast}\Gamma^{a a_1\ldots a_s}_{\mu\mu_1\ldots\mu_s}$ are
one-particle-irreducible amplitudes for the ultrasoft fields in the
thermal equilibrium or more concisely -- {\it ultrasoft amplitudes} (USA)
introduced by Blaizot and Iancu \cite{blaizot3}. USA represent the
generalization of usual HTL-amplitudes \cite{braaten} by including the
effects of the collisions among the thermal particles, in this case among
the hard transverse gluons.

Furthermore the classical parton current $j_{Q\mu}^a$ is given by expression
\cite{markov1}
\begin{equation}
j_{Q\mu}^{a}[A](p)=j_{Q\mu}^{(0)a}(p)+
\sum_{s=1}^{\infty} j_{Q\mu}^{(s)a}(A,\ldots,A),
\label{eq:3y}
\end{equation}
where $j_{Q\mu}^{(0)a}(p)=
g/(2\pi)^3Q^{a}_0\,\check{v}_{\mu}\delta(\check{v}\cdot p)$ is the initial
current of the energetic parton, and
\begin{equation}
j_{Q\mu}^{(s)a}(A,\ldots,A)= \check{v}_{\mu}\sum_{s=1}^{\infty}
\frac{g^{s+1}}{(2\pi)^3}\int\!\frac{1}
{(\check{v}\cdot (p_1 +\ldots +p_s))
(\check{v}\cdot (p_2 + \ldots + p_s))\ldots (\check{v}\cdot p_s)}
\label{eq:3u}
\end{equation}
\[
\times (\check{v}\cdot A^{a_1}(p_1))\ldots (\check{v}\cdot A^{a_s}(p_s))
\delta\biggl(\check{v}\cdot
\biggl(p-\sum_{i=1}^{s}p_i\biggr)\biggr)\prod_{i=1}^s dp_i
\,(T^{a_1}\ldots T^{a_s})^{ab}Q_0^b.
\]

Finally, the fluctuation current $\zeta_{\mu}^a$ functionally depending on
both the gauge field and the noise term (by virtue of equation for
fluctuating piece (\ref{eq:2u})) has a following structure
\begin{equation}
\zeta_{\mu}^{a}[A,\nu](p)=\zeta_{\mu}^{(0)a}[\nu](p)+
\sum_{s=1}^{\infty} \zeta_{\mu}^{(s)a}[A,\nu](p).
\label{eq:3i}
\end{equation}
Here,
\begin{equation}
\zeta_{\mu}^{(0)a}[\nu](p)\equiv\zeta_{\mu}^{(0)a}(p) =
im_D^2\!\int\frac{d\Omega_{\bf v}}{4\pi}
\int\frac{d\Omega_{{\bf v}_1}}{4\pi}\,v_{\mu}
\Bigl\langle v\Big|\,\frac{1}{v\cdot p +i\hat{\rm C}}
\,\Big|v_1\!\Bigr\rangle\,\nu^a(p,{\bf v}_1)
\label{eq:3o}
\end{equation}
is the fluctuation color current generated only by Langevin source on the
right-hand side of the Boltzmann equation (\ref{eq:2q}) and
\begin{equation}
\zeta_{\mu}^{(s)a}[A,\nu](p)\equiv
\zeta_{\mu}^{(s)a}(p) =
im_D^2\,g^s\!\!\int\frac{d\Omega_{\bf v}}{4\pi}\!
\int\prod\limits_{i=1}^{s+1}\frac{d\Omega_{{\bf v}_i}}{4\pi}
\label{eq:3p}
\end{equation}
\[
\times v_{\mu}\,
\Bigl\langle v\Big|\,\frac{1}{v\cdot p +i\hat{\rm C}}
\,\Big|v_1\Bigr\rangle\, v_1^{\mu_1}
\Bigl\langle v_1\Big|\,\frac{1}{v_1\cdot (p-p_1) +i\hat{\rm C}}
\,\Big|v_2\Bigr\rangle\,v_2^{\mu_2}\ldots
\]
\[
\ldots
v_{s-1}^{\mu_{s-1}}
\Bigl\langle v_{s-1}\Big|\,
\frac{1}{v_{s-1}\cdot (p-p_1-\ldots -p_{s-1}) +i\hat{\rm C}}
\,\Big|v_s\Bigr\rangle\, v_s^{\mu_s}\,
\Bigl\langle v_{s}\Big|\,
\frac{1}{v_{s}\cdot p_{s+1} +i\hat{\rm C}}
\,\Big|v_{s+1}\Bigr\rangle
\]
\[
\times
A^{a_1\mu_1}(p_{1})A^{a_2\mu_2}(p_{2})\ldots A^{a_s\mu_s}(p_{s})
\nu^b(p_{s+1},{\bf v}_{s+1})
{\delta}^{(4)}\!\biggl(p - \sum_{i=1}^{s+1}p_{i}\biggr)\prod_{i=1}^{s+1}dp_{i}.
\]

The next step is a derivation of a formal solution of basic field
equation (\ref{eq:3q})
by the approximation scheme method. Discarding the nonlinear terms in
$A_{\mu}$ and product $A_{\mu} \nu$ on the right-hand side of Eq.\,(\ref{eq:3q}),
we obtain in the first approximation
\[
^{\ast}{\cal D}_{\mu \nu}^{-1}(p) A^{a\nu}(p) =
-j_{Q\mu}^{(0)a}(p)-\zeta_{\mu}^{(0)a}(p),
\]
where $\zeta_{\mu}^{(0)a}(p)$ is defined by
Eq.\,(\ref{eq:3o}). The general solution of this equation is
\[
A_{\mu}^a(p) = A_{\mu}^{(0)a}(p) -\,^{\ast}{\cal D}_{\mu\nu}(p)
\left\{j_Q^{(0)a \nu} (p) + \zeta^{(0)a \nu} (p)\right\}.
\]
Here, $A_{\mu}^{(0)a} (p)$ is a solution of homogeneous equation
(\ref{eq:3q}) (free field), and the last term on the right-hand side
represents the field induced by a high-energy color particle in medium and
the noise term. From the physical point of view it is evident that
at the ultrasoft momentum scale, where the effects of the collisions among
the hard thermal particles are essential, free plasma waves in the system
are to be absent. In other words, the dispersion equation
${\rm det}(^{\ast}{\cal D}_{\mu\nu}^{-1}(p)) = 0$ with inverse propagator
(\ref{eq:3e}) at the ultrasoft momentum scale defines a very significant
damping rate of plasma waves preventing their existence and therefore one
can set $A_{\mu}^{(0)a}(p) \equiv 0$.
Thus the interaction field $A_{\mu}(p)$ defined as a
solution of Eq.\,(\ref{eq:3q}) with current (\ref{eq:3w}) will be represented
in the form of a functional expansion in powers of the noise term
$\nu (p, {\bf v})$ (Guerin, \cite{guerin2}) and color charge $Q_0^a$.

We can write out a more general structure of the interacting gauge field
as follows:
\begin{equation}
A_{\mu}^a(p) = - ^{\ast}{\cal D}_{\mu\nu}(p)J^{({\rm tot})a\nu}
[\nu, Q_0] (p),
\label{eq:3a}
\end{equation}
where the total effective current\footnote{The effective current will be
designated by capital letter $J_{\mu}^a$.} is
\begin{equation}
J_{\mu}^{({\rm tot})a}[\nu,Q_0](p) =
\sum_{n=0}^{\infty} J_{Q \mu}^{(n)a} [\nu] (p) +
\sum_{n=1}^{\infty} J_{\mu}^{(n)a} [\nu] (p).
\label{eq:3s}
\end{equation}
Here, the terms $J_{Q \mu}^{(n)a}[\nu](p)$ and $J_{\mu}^{(n)a}[\nu](p)$ are
proportional to the $n$th power of $\nu$; $J_{Q \mu}^{(0)a} (p)
\equiv j_{Q \mu}^{(0)a} (p)$ is initial current of the energetic
parton and $J_{\mu}^{(1)}[\nu](p)\equiv\zeta_{\mu}^{(0)a}
[\nu](p)$ is initial fluctuation current. The last sum on the
right-hand side of Eq.\,(\ref{eq:3s}) is independent of color charge
$Q_0$ and describes the processes of self-interaction of ultrasoft
plasma fluctuations. The first sum contains relevant information
on the interaction of the energetic parton with ultrasoft
fluctuations. Below we will calculate the first
correction $J_{Q\mu}^{(1)a}$ to the initial current
$J_{Q\mu}^{(0)a} =
g/(2\pi)^3Q_0^a\check{v}_{\mu}\delta(\check{v}\cdot p)$.
For this purpose we need the relations
\begin{equation}
A_{\mu}^a(p)|_{\nu=Q_0=0}=0, \quad
\left.\frac{\delta A^{a\mu}(p)}{\delta Q_0^b}\right|_{\nu=Q_0=0}
=-\,\frac{g}{(2\pi)^3}\,\delta^{ab}\,^{\ast}{\cal D}^{\mu\nu}(p)
\check{v}_{\nu}\delta (\check{v}\cdot p),
\label{eq:3d}
\end{equation}
\[
\left.\frac{\delta A^{a\mu}(p)}{\delta \nu^b(p^{\prime},{\bf v}^{\prime})}
\right|_{\nu=Q_0=0}\!\!=\!
-^{\ast}{\cal D}^{\mu\nu}(p)
\frac{\delta\zeta_{\mu}^{(0)a}[\nu](p)}
{\delta\nu^b(p^{\prime},{\bf v}^{\prime})}
= im_D^2\delta^{ab}\!\,^{\ast}{\cal D}^{\mu\nu}(p)\!
\!\int\!\frac{d\Omega_{\bf v}}{4\pi}
v_{\nu}\,\Bigl\langle v\Big|\,\frac{1}{v\cdot p +i\hat{\rm C}}
\,\Big|v^{\prime}\Bigr\rangle\delta(p-p^{\prime})
\]
following from general formula (\ref{eq:3a}).

\begin{figure}[hbtp]
\begin{center}
\includegraphics*[height=6cm]{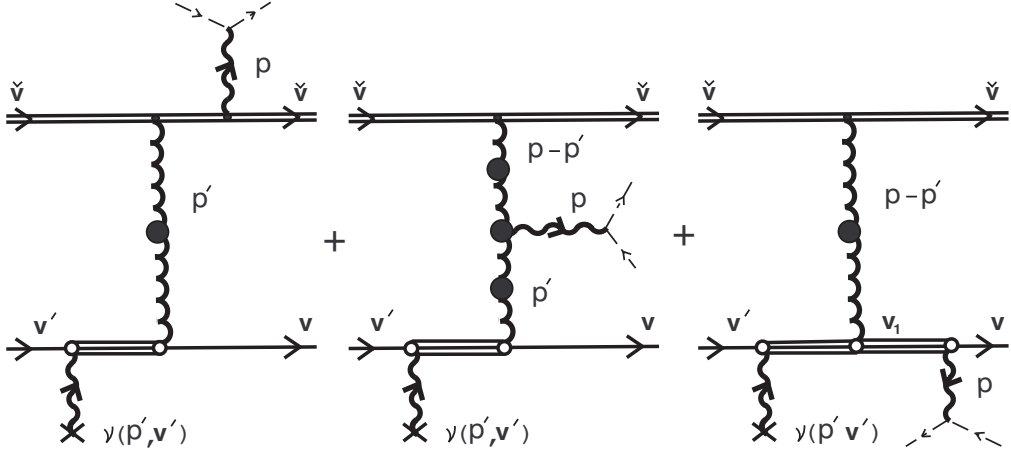}
\end{center}
\caption{\small The process of the scattering of the energetic
parton off ultrasoft plasma fluctuations in the linear
approximation over both noise $\nu(p^{\prime},{\bf v}^{\prime})$
and initial color charge of parton $Q_0$. The blob on the waveline
stands for the ultrasoft-gluon propagator and 3-gluon
ultrasoft vertex. The double line denotes energetic parton with
velocity $\check{\bf v}$. Triple line denotes collision-resummed
propagator $\Bigl\langle v\Big|\,\frac{1}{v\cdot p^{\prime}
+i\hat{\rm C}} \,\Big|v^{\prime}\Bigr\rangle$, and cross denotes
the noise source.} \label{fig1}
\end{figure}

According to the general approach to computation of effective currents
\cite{markov1, markov2}
the expression for the first correction $J_{Q\mu}^{(1)a}$ in the leading order
in the coupling constant (i.e. a linear part over color charge $Q_0$) is
defined by the following expression:
\small{
\[
J_{Q\mu}^{(1)a}[\nu](p)\!=\!
Q_0^c\!\!\int\!\!dp^{\prime}\!\!\int\!\!
\frac{d\Omega_{{\bf v}^{\prime}}}{4\pi}\,
\nu^b(p^{\prime},{\bf v}^{\prime})\!\!
\left.\left[
\frac{\delta^2j^{\,{\rm ind}(2)a}_{\mu}[A](p)}
{\delta\nu^b(p^{\prime},{\bf v}^{\prime})\delta Q_0^c}
+\frac{\delta^2j^{(1)a}_{Q\mu}[A](p)}{\delta\nu^b(p^{\prime},{\bf v}^{\prime})
\delta Q_0^c} +
\frac{\delta^2\zeta^{(1)a}_{\mu}[A,\nu](p)}
{\delta\nu^b(p^{\prime},{\bf v}^{\prime})\delta Q_0^c}
\right]\right|_{\nu=Q_0=0}.
\]
}\normalsize
By using explicit expressions for the currents $j^{\,{\rm ind}(2)a}_{\mu},\,
j^{(1)a}_{Q\mu}$ and $\zeta^{(1)a}_{\mu}$ (Eqs.\,(\ref{eq:3t}), (\ref{eq:3u})
and (\ref{eq:3p})) and relations (\ref{eq:3d}), after simple calculations
we arrive at the expression for the desired current correction
\begin{equation}
J_{Q\mu}^{(1)a}[\nu](p)=
-i(T^c)^{ab}Q_0^c\,\biggl(\frac{\alpha_s}{2\pi^2}\biggr)m_D^2\!
\int\!dp^{\prime}\!\int\!\frac{d\Omega_{{\bf v}^{\prime}}}{4\pi}\,
{\cal R}_{\mu}(p,p^{\prime};\check{v},v^{\prime})
\nu^b(p^{\prime},{\bf v}^{\prime})\,\delta(\check{v}\cdot(p-p^{\prime})),
\label{eq:3f}
\end{equation}
where $\alpha_s=g^2/4\pi$, and
\begin{equation}
{\cal R}_{\mu}(p,p^{\prime};\check{v},v^{\prime}) =
-\frac{\check{v}_{\mu}}{(\check{v}\cdot p^{\prime})}
\int\!\frac{d\Omega_{\bf v}}{4\pi}\,\left(\check{v}_{\nu}
\!\,^{\ast}{\cal D}^{\nu\nu^{\prime}}(p^{\prime})v_{\nu^{\prime}}\right)\,
\Bigl\langle v\Big|\,\frac{1}{v\cdot p^{\prime} +i\hat{\rm C}}
\,\Big|v^{\prime}\Bigr\rangle
\label{eq:3g}
\end{equation}
\[
-\,\!\,^{\ast}\Gamma_{\mu\nu\lambda}(p,-p^{\prime},-p+p^{\prime})
\,^{\ast}{\cal D}^{\nu\nu^{\prime}}(p^{\prime})\!
\int\!\frac{d\Omega_{\bf v}}{4\pi}\,{v}_{\nu^{\prime}}
\Bigl\langle v\Big|\,\frac{1}{v\cdot p^{\prime} +i\hat{\rm C}}
\,\Big|v^{\prime}\Bigr\rangle
\,^{\ast}{\cal D}^{\lambda\lambda^{\prime}}(p-p^{\prime})
\check{v}_{\lambda^{\prime}}
\]
\[
+\,
\int\!\frac{d\Omega_{\bf v}}{4\pi}
\int\!\frac{d\Omega_{{\bf v}_1}}{4\pi}\,v_{\mu}
\Bigl\langle v\Big|\,\frac{1}{v\cdot p +i\hat{\rm C}}
\,\Big|v_1\Bigr\rangle\,v_{1\nu}\,
\Bigl\langle v_1\Big|\,\frac{1}{v_1\cdot p^{\prime} +i\hat{\rm C}}
\,\Big|v^{\prime}\Bigr\rangle
\,^{\ast}{\cal D}^{\nu\nu^{\prime}}(p-p^{\prime})
\check{v}_{\nu^{\prime}}.
\]
Remind that the three-gluon vertex $\!\,^{\ast}\Gamma_{\mu \mu_1 \mu_2}$ on the
right-hand side of Eq.\,(\ref{eq:3g}) is ultrasoft amplitude. Its explicit
form is defined in Ref.\,\cite{guerin1}. The diagrammatic interpretation of
different terms on the right-hand side of Eq.\,(\ref{eq:3g}) is presented in
Fig.\,\ref{fig1}.

We note also that the first two terms on the right-hand side of
Eq.\,(\ref{eq:3g}) can be presented in a more compact form
\begin{equation}
-\stackrel{\scriptscriptstyle{\,(1)}}{\displaystyle{K}}_{\mu\nu}
\!(\check{\bf v}\vert\,p,-p^{\prime})
\,^{\ast}{\cal D}^{\nu\nu^{\prime}}(p^{\prime})
\!\int\!\frac{d\Omega_{\bf v}}{4\pi}\,v_{\nu^{\prime}}\,
\Bigl\langle v\Big|\,\frac{1}{v\cdot p^{\prime} +i\hat{\rm C}}
\,\Big|v^{\prime}\Bigr\rangle,
\label{eq:3h}
\end{equation}
where the function
\[
\stackrel{\scriptscriptstyle{\,(1)}}{\displaystyle{K}}_{\mu\nu}
\!(\check{\bf v}\vert\,p,-p^{\prime})=
\frac{\check{v}_{\mu}\check{v}_{\nu}}{\check{v}\cdot p^{\prime}} +
\!\,^\ast\Gamma_{\mu\nu\lambda}(p,-p^{\prime},-p+p^{\prime})
\,^{\ast}{\cal D}^{\lambda\lambda^{\prime}}\!(p-p^{\prime})
\check{v}_{\lambda^{\prime}}
\]
was introduced in work \cite{markov1}. This function defines a
matrix element of the nonlinear damping process of soft
plasma excitations.
In the next section we consider the convolution
$\bar{u}_{\mu}(p){\cal R}^{\mu}(p,
p^{\prime};\check{v},v^{\prime})$, where $\bar{u}_{\mu}(p)$ is
the longitudinal projector in the covariant gauge (see Appendix).
Such a convolution
appears in the problem of the energy losses (Section 5). We prove
that this convolution is gauge-invariant at least in the classes
of temporal and covariant gauges in {\it a weak sense}, in spirit
of B\"odeker \cite{bodeker1}.

The general expression for a current correction of second order
over $\nu$ in linear approximation over the color charge $Q_0$ has the
following form:
\[
J_{Q\mu}^{(2)a}[\nu](p)=
Q_0^d\,g\biggl(\frac{\alpha_s}{2\pi^2}\biggr)m_D^4\!
\int\!dp^{\prime}\!\int\!dp^{\prime\prime}
\!\int\!\frac{d\Omega_{{\bf v}^{\prime}}}{4\pi}
\!\int\!\frac{d\Omega_{{\bf v}^{\prime\prime}}}{4\pi}\,
{\cal R}_{\mu}^{abcd}(p,p^{\prime},p^{\prime\prime};
\check{v},v^{\prime},v^{\prime\prime})
\]
\begin{equation}
\times\,\Bigl\{
\nu^b(p^{\prime},{\bf v}^{\prime})
\nu^c(p^{\prime\prime},{\bf v}^{\prime\prime})\,-
\ll\!\nu^b(p^{\prime},{\bf v}^{\prime})
\nu^c(p^{\prime\prime},{\bf v}^{\prime\prime})
\!\gg\Bigr\}
\,\delta(\check{v}\cdot(p-p^{\prime}-p^{\prime\prime})),
\label{eq:3j}
\end{equation}
where
\begin{equation}
{\cal R}_{\mu}^{abcd}(p,p^{\prime},p^{\prime\prime};
\check{v},v^{\prime},v^{\prime\prime})=
f^{abe}f^{ecd}
{\cal R}_{\mu}(p,p^{\prime},p^{\prime\prime};
\check{v},v^{\prime},v^{\prime\prime})
+f^{ace}f^{ebd}
{\cal R}_{\mu}(p,p^{\prime\prime},p^{\prime};
\check{v},v^{\prime\prime},v^{\prime}).
\label{eq:3k}
\end{equation}

We will not to write out an explicit expression for function
${\cal R}_{\mu}(p,p^{\prime},p^{\prime\prime};
\check{v},v^{\prime},v^{\prime\prime})$ by virtue of its excessive awkwardness.
On the right-hand side of Eq.\,(\ref{eq:3j}) we insert a term with
$\ll\!\nu\nu\!\gg$ to maintain stochasticity of the gauge field (\ref{eq:3a})
(condition $\ll\!A_{\mu}\!\gg = 0$).

\begin{figure}[hbtp]
\begin{center}
\includegraphics*[height=5cm]{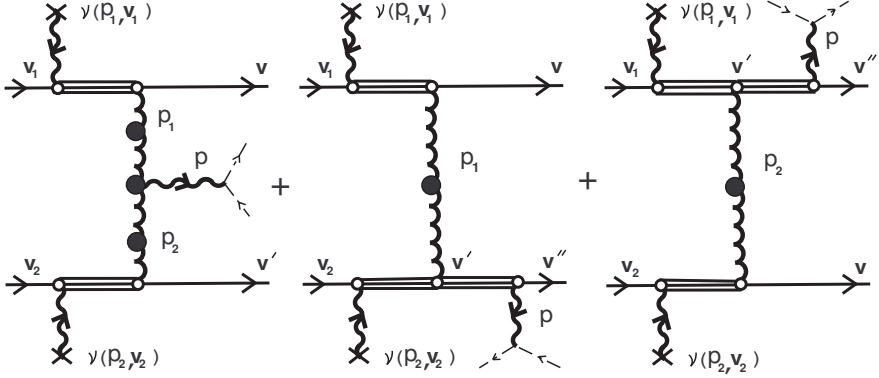}
\end{center}
\caption{\small The simplest process of nonlinear self-interaction of
ultrasoft gluon fluctuations.}
\label{fig2}
\end{figure}

In closing this section we give also an expression for current
$J_{\mu}^{(2)a}[\nu](p)$ in the second
sum of expansion of the effective current (\ref{eq:3s}). This current defines
the simplest process of the nonlinear interaction of the thermal ultrasoft
fluctuations produced by the noise term $\nu$ among themselves. Computing
a second order variation in $\nu^{a}(p,{\bf v})$ of initial expression for
the total current (\ref{eq:3w}) and taking into account (\ref{eq:3d})
we obtain
\[
J_{\mu}^{(2)a}[\nu](p)=
-g m_D^4(T^a)^{a_1a_2}\!
\int\!\!dp_1\!\int\!\!dp_1\!\int\!\frac{d\Omega_{{\bf v}_1}}{4\pi}\!
\int\!\frac{d\Omega_{{\bf v}_2}}{4\pi}\,
{\cal T}_{\mu}(p,-p_1,-p_2;v_1,v_2)
\]
\[
\times\,
\nu^{a_1}(p_1,{\bf v}_1)\nu^{a_2}(p_2,{\bf v}_2)
\,\delta(p-p_1-p_2),
\]
where
\[
{\cal T}_{\mu}(p,-p_1,-p_2;v_1,v_2) \equiv
\,\!\,^{\ast}\Gamma_{\mu\mu_1\mu_2}(p,-p_1,-p_2)
\,^{\ast}{\cal D}^{\mu_1\mu_1^{\prime}}(p_1)\!
\left\{
\int\!\frac{d\Omega_{\bf v}}{4\pi}\,{v}_{\mu_1^{\prime}}
\Bigl\langle v\Big|\,\frac{1}{v\cdot p_1 +i\hat{\rm C}}
\,\Big|v_1\Bigr\rangle\right\}
\]
\[
\times
\,^{\ast}{\cal D}^{\mu_2\mu_2^{\prime}}(p_2)\!
\left\{
\int\!\frac{d\Omega_{{\bf v}^{\prime}}}{4\pi}\,{v}_{\mu_2^{\prime}}^{\prime}
\Bigl\langle v^{\prime}\Big|\,\frac{1}{v^{\prime}\cdot p_2 +i\hat{\rm C}}
\,\Big|v_2\Bigr\rangle\!\right\}
+
\int\!\frac{d\Omega_{\bf v}}{4\pi}\!
\int\!\frac{d\Omega_{{\bf v}^{\prime}}}{4\pi}\!
\int\!\frac{d\Omega_{{\bf v}^{\prime\prime}}}{4\pi}
\,v_{\mu}
\Bigl\langle v\Big|\,\frac{1}{v\cdot p +i\hat{\rm C}}
\,\Big|v^{\prime}\Bigr\rangle
\]
\[
\times
\Biggl[
\,v^{\prime}_{\mu_1}\,
\Bigl\langle v^{\prime}\Big|\,\frac{1}{v^{\prime}\cdot p_2 +i\hat{\rm C}}
\,\Big|v_2\Bigr\rangle
\,^{\ast}{\cal D}^{\mu_1\mu_1^{\prime}}(p_1)
\,v^{\prime\prime}_{\mu^{\prime}_1}
\Bigl\langle v^{\prime\prime}\Big|\,
\frac{1}{v^{\prime\prime}\cdot p_1 +i\hat{\rm C}}
\,\Big|v_1\Bigr\rangle\,-
\]
\[
-\,v^{\prime}_{\mu_2}\,
\Bigl\langle v^{\prime}\Big|\,\frac{1}{v^{\prime}\cdot p_1 +i\hat{\rm C}}
\,\Big|v_1\Bigr\rangle
\,^{\ast}{\cal D}^{\mu_2\mu_2^{\prime}}(p_2)
\,v^{\prime\prime}_{\mu^{\prime}_2}
\Bigl\langle v^{\prime\prime}\Big|\,
\frac{1}{v^{\prime\prime}\cdot p_2 +i\hat{\rm C}}
\,\Big|v_2\Bigr\rangle
\Biggr].
\]
The diagrammatic interpretation of different contributions determining the
process of nonlinear self-interaction is presented on Fig.\,\ref{fig2}.

\section{\bf Gauge invariance of
$\bar{u}_{\mu}(p){\cal R}^{\mu}(p, p^{\prime};\check{v},v^{\prime})$}
\setcounter{equation}{0}

By using an explicit expression (\ref{eq:3g}) for the function
${\cal R}^{\mu}(p, p^{\prime};\check{v},v^{\prime})$ and the effective Ward
identity for the ultrasoft three-gluon amplitude \cite{blaizot3, guerin1},
we can represent the convolution $\bar{u}_{\mu}{\cal R}^{\mu}$ in the
following form
\[
\bar{u}_{\mu}(p){\cal R}^{\mu}(p, p^{\prime};\check{v},v^{\prime})
=p^2{\cal R}^{0}(p, p^{\prime};\check{v},v^{\prime})+
\omega\left[
\int\!\frac{d\Omega_{\bf v}}{4\pi}\,\left(\check{v}_{\nu}
\!\,^{\ast}{\cal D}^{\nu \nu^{\prime}}(p^{\prime})v_{\nu^{\prime}}\right)
\Bigl\langle v\Big|\,\frac{1}{v\cdot p^{\prime} +i\hat{\rm C}}
\,\Big|v^{\prime}\Bigr\rangle\right.
\]
\[
+\,
\Biggl\{
\biggl(
\delta_{\lambda}^{\nu^{\prime}}-
\frac{p^{\prime\nu^{\prime}}p^{\prime}_{\lambda}}{p^{\prime 2}}\biggr)
\,^{\ast}{\cal D}^{\lambda\lambda^{\prime}}(p-p^{\prime})\check{v}_{\lambda^{\prime}}-
\biggl(
\delta_{\nu}^{\lambda^{\prime}}-
\frac{(p-p^{\prime})^{\lambda^{\prime}}(p-p^{\prime})_{\nu}}
{(p-p^{\prime})^2}\biggr)\check{v}_{\lambda^{\prime}}
\!\,^{\ast}{\cal D}^{\nu\nu^{\prime}}(p^{\prime})
\Biggr\}\times
\]
\begin{equation}
\times\!
\int\!\frac{d\Omega_{\bf v}}{4\pi}\,{v}_{\nu^{\prime}}
\Bigl\langle v\Big|\,\frac{1}{v\cdot p^{\prime} +i\hat{\rm C}}
\,\Big|v^{\prime}\Bigr\rangle
\hspace{8cm}
\label{eq:4q}
\end{equation}
\[
-\,\left.
\int\!\frac{d\Omega_{\bf v}}{4\pi}
\int\!\frac{d\Omega_{{\bf v}_1}}{4\pi}\,(v\cdot p)\,
\Bigl\langle v\Big|\,\frac{1}{v\cdot p +i\hat{\rm C}}
\,\Big|v_1\Bigr\rangle\,v_{1\nu}\,
\Bigl\langle v_1\Big|\,\frac{1}{v_1\cdot p^{\prime} +i\hat{\rm C}}
\,\Big|v^{\prime}\Bigr\rangle\,
\!\,^{\ast}{\cal D}^{\nu\nu^{\prime}}(p-p^{\prime})
\check{v}_{\nu^{\prime}}\right].
\]
In deriving (\ref{eq:4q}) we consider that $\check{v}\cdot(p-p^{\prime})=0$
by virtue of $\delta$-function in integrand of Eq.\,(\ref{eq:3f}). Below we
will show that on the right-hand side of Eq.\,(\ref{eq:4q}) all terms
proportional to $\omega$ either equal to zero or mutually cancel out.

In projector
\[
\biggl(
\delta_{\nu}^{\lambda^{\prime}}-
\frac{(p-p^{\prime})^{\lambda^{\prime}}(p-p^{\prime})_{\nu}}
{(p-p^{\prime})^2}\biggr)
\]
in convolution with $\check{v}_{\lambda^{\prime}}$ the last term
vanishes by virtue of the above-mentioned constraint
$\check{v}\cdot(p-p^{\prime})=0$, and contribution of
$\delta_{\nu}^{\lambda^{\prime}}$ cancels with the second term on the
right-hand side of Eq.\,(\ref{eq:4q}).

Furthermore we transform the last term on the right-hand side of
Eq.\,(\ref{eq:4q}). Here, we use a relation
\begin{equation}
\int\!\frac{d\Omega_{\bf v}}{4\pi}\,(v\cdot p)\,
\Bigl\langle v\Big|\,\frac{1}{v\cdot p +i\hat{\rm C}}
\,\Big|\,v_1\!\Bigr\rangle=
\int\!\frac{d\Omega_{\bf v}}{4\pi}\,
\delta^{(S^2)}({\bf v}-{\bf v}_1)=1,
\label{eq:4w}
\end{equation}
which is true by virtue of the property:
\begin{equation}
\int\!\frac{d\Omega_{\bf v}}{4\pi}\,
\Bigl\langle v\Big|\,\hat{\rm C}\,\Big|v_1\!\Bigr\rangle=
\int\!\frac{d\Omega_{\bf v}}{4\pi}\;
{\cal C}({\bf v},{\bf v}_1)=0.
\label{eq:4e}
\end{equation}
Taking into account the above-mentioned relation for the last term in
Eq.\,(\ref{eq:4q}) after replacement $v_1\rightarrow v$, we can
present this term as follows:
\[
\int\!\frac{d\Omega_{\bf v}}{4\pi}\,v_{\nu}\,
\Bigl\langle v\Big|\,\frac{1}{v\cdot p^{\prime} +i\hat{\rm C}}
\,\Big|v^{\prime}\Bigr\rangle
\,^{\ast}{\cal D}^{\nu\nu^{\prime}}(p-p^{\prime})
\check{v}_{\nu^{\prime}}
\]
that exactly cancels out with the first term from a projector
\[
\biggl(\delta_{\lambda}^{\nu^{\prime}}-
\frac{p^{\prime\nu^{\prime}}p^{\prime}_{\lambda}}{p^{\prime 2}}\biggr).
\]
Thus on the right-hand side of Eq.\,(\ref{eq:4q}), besides a first term
$p^2{\cal R}^0$ different from zero, the following term is left
\begin{equation}
-\,\frac{p^{\prime}_{\lambda}}{p^{\prime 2}}
\,^{\ast}{\cal D}^{\lambda\lambda^{\prime}}(p-p^{\prime})
\check{v}_{\lambda^{\prime}}
\!\int\!\frac{d\Omega_{\bf v}}{4\pi}\,(v\cdot p^{\prime})
\Bigl\langle v\Big|\,\frac{1}{v\cdot p^{\prime} +i\hat{\rm C}}
\,\Big|v^{\prime}\Bigr\rangle
\equiv-\,\frac{p^{\prime}_{\lambda}}{p^{\prime 2}}
\,^{\ast}{\cal D}^{\lambda\lambda^{\prime}}(p-p^{\prime})
\check{v}_{\lambda^{\prime}}.
\label{eq:4r}
\end{equation}
Here, on the right-hand side we take into account relation (\ref{eq:4w}).
Thus this term is independent of unit vector ${\bf v}^{\prime}$.
The consequence of this important property
will be the fact that in the expression for current
$J_{Q\mu}^{(1)a}[\nu](p)$ (\ref{eq:3f}) the solid integration
$\int\!d\Omega_{{\bf v}^{\prime}}/4\pi$
appears only in combination
\[
\int\!\frac{d\Omega_{{\bf v}^{\prime}}}{4\pi}\,
\nu^b(p^{\prime},{\bf v}^{\prime}).
\]
Furthermore in subsequent discussion we will follow reasoning of B\"odeker
\cite{bodeker1}.

In the expression for the energy losses of energetic color parton
(Eq.\,(\ref{eq:5q})) the noise
$\nu^b(p^{\prime},{\bf v}^{\prime})$ enters only in the form of the 2-point
function: $\ll\!\nu\nu\!\gg$. However the correlation function of
$\nu^a(p,{\bf v})$ and $\int\!d\Omega_{{\bf v}^{\prime}}/4\pi\,
\nu^b(p^{\prime},{\bf v}^{\prime})$ vanishes
\[
\ll\! \nu^a(p,{\bf v}) \!\int\!\frac{d\Omega_{{\bf v}^{\prime}}}{4\pi}\,
\nu^b(p^{\prime},{\bf v}^{\prime})
\!\gg\,=0
\]
due to (\ref{eq:2r}) and (\ref{eq:4e}). Thus we can set
$\int\!d\Omega_{{\bf v}^{\prime}}/4\pi\,\nu^b(p^{\prime},{\bf
v}^{\prime}) \stackrel{\rm w}{=}0$, where the symbol `w' denotes
equality in a weak sense, so that the term (\ref{eq:4r}) can be
set equal to zero in a weak sense.

Taking into account the above-mentioned, we can write the convolution
(\ref{eq:4q}) in the form
\begin{equation}
\bar{u}_{\mu}(p){\cal R}^{\mu}(p, p^{\prime};\check{v},v^{\prime})
\stackrel{\rm w}{=}
p^2{\cal R}^{0}(p, p^{\prime};\check{v},v^{\prime}).
\label{eq:4t}
\end{equation}
Furthermore we consider the convolution $\bar{u}_{\mu}{\cal R}^{\mu}$ in the
temporal gauge. For this purpose we perform replacements (see Appendix)
\[
\bar{u}_{\mu}(p)\rightarrow
\tilde{u}_{\mu}(p)=\frac{p^2}{(p\cdot u)}(p_{\mu} - u_{\mu} (p\cdot u)),
\]
\[
\,^{\ast}{\cal D}_{\mu\nu}(p)\rightarrow
\!\,^{\ast}\tilde{\cal D}_{\mu \nu}(p)= -
P_{\mu \nu}(p) \,^{\ast}\!\Delta^t(p) -
\tilde{Q}_{\mu \nu}(p) \,^{\ast}\!\Delta^l(p) +
\xi_0\frac{p^2}{(p\cdot u)^2}\,D_{\mu \nu}(p),
\]
where $\xi_0$ is a gauge parameter fixing the temporal gauge. Then after
analogous computations we come to the expression
\begin{equation}
\tilde{u}_{\mu}(p){\cal R}^{\mu}(p, p^{\prime};\check{v},v^{\prime})
\stackrel{\rm w}{=}
p^2{\cal R}^{0}(p, p^{\prime};\check{v},v^{\prime}).
\label{eq:4y}
\end{equation}
All the terms on the right-hand side of Eq.\,(\ref{eq:4y}) (as well as
Eq.\,(\ref{eq:4t})) including gauge
parameter vanish. Thus as follows from equalities (\ref{eq:4t}) and
(\ref{eq:4y}), at least in the classes of the temporal and covariant gauges
the function $\bar{u}_{\mu}{\cal R}^{\mu}$ is gauge-invariant in a weak
sense.

\section{Energy losses of energetic parton induced by scattering off
ultrasoft gluon fluctuations}
\setcounter{equation}{0}

As an application of the formalism developed in previous sections we consider
the problem of calculation of the energy losses of the energetic parton induced
by scattering process off ultrasoft boson excitations of the medium. As initial
expression for the energy loss a classical expression for parton
energy loss per unit length (Eq.\,(7.1) in \cite{markov1}) with replacement
of the expectation value $<\!\cdot\!>$ over statistical ensemble
by averaging $\ll\!\cdot\!\gg$ over the random white noise) is used
\begin{equation}
\Biggl(-\frac{dE}{dx}\Biggr)_{\!{\rm ul.soft}}\! =
\frac{1}{\vert\check{\bf v}\vert}
\lim\limits_{\tau\rightarrow\infty}
\frac{1}{\tau}\!\int\limits_{-\tau/2}^{\tau/2}\!\int
\!d{\bf x}\,dt\!\int\!\!dQ_0\,{\rm Re}\!
\ll\!{\bf J}^a_Q ({\bf x},t)\cdot {\bf E}^a_Q ({\bf x},t)\!\gg
\label{eq:5q}
\end{equation}
\[
=-\frac{1}{\vert\check{\bf v}\vert}\,
\lim\limits_{\tau\rightarrow\infty}
\frac{(2\pi)^4}{\tau}
\int\!\!d{\bf p}\,d\omega\!\int\!\!dQ_0\;
\omega\,
{\rm Im}\!\ll\!
J^{a}_{Q\mu}(p)
\,^{\ast}\tilde{\cal D}^{\mu\nu}(p)
J^{\ast a}_{Q\nu}(p)\!\gg.
\]
Here, the color integration is
\[
\int\!dQ_0=\int\!\prod_{a=1}^{d_A}dQ_0^a\,\delta(Q_0^aQ_0^a-C_2),\quad
d_A = N_c^2 - 1,
\]
where $C_2$ is the second order Casimir ($C_2=C_A$ for energetic gluon and
$C_2=C_F$ for energetic quark). In the last line of Eq.\,(\ref{eq:5q})
$J^a_{Q\mu}(p)\equiv J_{Q\mu}^{a}[\nu,Q_0](p) \cong
\sum_{n=0}^{\infty}J_{Q \mu}^{(n)a}[\nu](p)Q_0^b$ is the effective color
current of hard parton and $\!\,^{\ast}\tilde{\cal D}^{\mu\nu}(p)$ is a
gluon propagator (in the temporal gauge) at the ultrasoft momentum scale
\begin{equation}
\,^{\ast}\tilde{\cal D}^{ij}(p) =
\left(\frac{p^2}{\omega^2}\right)
\frac{{\rm p}^i{\rm p}^j}{{\bf p}^2}
\,^{\ast}\!\Delta^{l}(p) +
\left(\delta^{ij} - \frac{{\rm p}^i{\rm p}^j}{{\bf p}^2}\right)
\!\!\,^{\ast}\!\Delta^{t}(p),\,\,
\,^{\ast}\tilde{\cal D}^{i0}(p) =
\,^{\ast}\tilde{\cal D}^{0i}(p) =
\,^{\ast}\tilde{\cal D}^{00}(p) = 0.
\label{eq:5w}
\end{equation}

First of all, we write out the expression for the energy loss connected
with initial current $J_{Q\mu}^{(0)a}(p)=
g/(2\pi)^3\delta^{ab}\,\check{v}_{\mu}\delta(\check{v}\cdot p)Q_0^a.$
Substituting this current into (\ref{eq:5q}), taking into account the structure
of propagator (\ref{eq:5w}) and rules
\[
\int\!dQ_0\,Q_0^aQ_0^b = \frac{C_2}{d_A}\,\delta^{ab},\quad
\Bigr[\delta(v\cdot p)\Bigl]^2 =
\frac{1}{2\pi}\,\tau\delta(v\cdot p),
\]
we obtain
\begin{equation}
\left(\!-\frac{dE^{(0)}}{dx}\right)_{\!{\rm ul.soft}}\!
= -\frac{1}{\vert\check{\bf v}\vert}
\Biggl(\frac{N_c\alpha_s}{2{\pi}^2}\Biggr)\!
\Biggl(\frac{C_2}{N_c}\Biggr)\!
\int\! dp\,\frac{\omega}{{\bf p}^2}
\biggl\{{\rm Im}\,(p^2\!\,^{\ast}{\!\Delta}^l(p)) +
(\check{\bf v}\times {\bf p})^2
{\rm Im}\,(^{\ast}{\!\Delta}^t(p))\biggr\}
\delta(\check{v}\cdot p).
\label{eq:5r}
\end{equation}
This expression defines the polarization losses of energetic parton
connected with large distance collisions \cite{thoma}.
However unlike \cite{thoma} here integrand (and in particular scalar
propagators
$\!\,^{\ast}\!\Delta^{t,\,l}(p)=1/(p^2 - \Pi^{t,\,l}(p))\,$)
is defined at the ultrasoft momentum scale.

Now we define the expression for the energy losses produced by the following
term in the expansion of the effective current:
$J_{Q\mu}^{(1)a}[\nu,Q_0](p) \cong J_{Q \mu}^{(1)ab}[\nu](p)\,Q_0^b$
(Eq.\,(\ref{eq:3f}), (\ref{eq:3g})).
For this purpose we substitute the current (\ref{eq:3f}) into
general expression for the energy losses (\ref{eq:5q}). For the sake of
simplification we keep only a longitudinal part in propagator
$\!\,^{\ast}\tilde{\cal D}^{\mu\nu}(p)$
\begin{equation}
\,^{\ast}\tilde{\cal D}^{\mu\nu}(p) \rightarrow
-\,\tilde{Q}^{\mu\nu}(p)\,^{\ast}{\!\Delta}^{l}(p) \equiv
-\,\frac{\tilde{u}^{\mu}(p)\tilde{u}^{\nu} (p)}{\bar{u}^2(p)}
\;^{\ast}{\!\Delta}^{l}(p).
\label{eq:5t}
\end{equation}
Integrating over initial color charge, averaging with respect to the noise
term and taking into account weak equality (\ref{eq:4y}), we obtain the
following (after (\ref{eq:5r})) contribution to the energy
loss of the energetic parton
\[
\left(\!-\frac{dE^{(1)\,l}}{dx}\right)_{\!{\rm ul.soft}}\! =
-\frac{(2\pi)^7}{\vert\check{\bf v}\vert}\,2Tm_D^2
\Biggl(\frac{N_c\alpha_s}{2{\pi}^2}\Biggr)^{\!2}\!
\Biggl(\frac{C_2}{N_c}\Biggr)\!
\int\!\!dp\!\int\!\!dp^{\prime}
\,\biggl(\frac{\omega}{{\bf p}^2}\biggr)
\,{\rm Im}\Bigl(p^2\!\,^{\ast}{\!\Delta}^l(p)\Bigr)
\,\delta(\check{v}\cdot(p-p^{\prime}))
\]
\begin{equation}
\times\!\int\!\frac{d\Omega_{{\bf v}}}{4\pi}
\!\int\!\frac{d\Omega_{{\bf v}^{\prime}}}{4\pi}
\,\Bigl[
{\cal R}_{0}(p,p^{\prime};\check{v},v)
{\cal C}({\bf v},{\bf v}^{\prime})
{\cal R}^{\ast}_0(p,p^{\prime};\check{v},v^{\prime})
\Bigr].
\label{eq:5y}
\end{equation}
Here, the function ${\cal R}_{0}(p,p^{\prime};\check{v},v^{\prime})$ is defined
by Eq.\,(\ref{eq:3g}), and ${\cal C}({\bf v},{\bf v}^{\prime})$
is a collision kernel defined by Eq.\,(\ref{eq:2e}). The contribution to the
energy loss (\ref{eq:5y}) of the term (\ref{eq:3h}) can be presented in the
following form:
\begin{equation}
-\frac{(2\pi)^7}{\vert\check{\bf v}\vert}\,m_D^2
\Biggl(\frac{N_c\alpha_s}{2{\pi}^2}\Biggr)^{\!2}\!
\Biggl(\frac{C_2}{N_c}\Biggr)\!
\int\!\!dp\!\int\!\!dp^{\prime}
\,\biggl(\frac{\omega}{{\bf p}^2}\biggr)
\,{\rm Im}\Bigl(p^2\!\,^{\ast}{\!\Delta}^l(p)\Bigr)
\delta(\check{v}\cdot(p-p^{\prime}))
\label{eq:5u}
\end{equation}
\[
\times\,\Bigl[
\stackrel{\scriptscriptstyle{\,(1)}}{\displaystyle{K}}_{0\nu}
\!(\check{\bf v}\vert\,p,-p^{\prime})
\,^{\ast}\tilde{\cal D}^{\nu\nu^{\prime}}\!(p^{\prime})
\biggl(-\frac{2T}{\omega^{\prime}}\biggr)
\,{\rm Im}\Pi_{\nu^{\prime}\lambda^{\prime}}^{(R)}(p^{\prime})\!
\stackrel{\scriptscriptstyle{\,(1)}}{\displaystyle{K}^{\ast}}_{\!\!0\lambda}
\!(\check{\bf v}\vert\,p,-p^{\prime})
\,^{\ast}\tilde{\cal D}^{\ast\lambda\lambda^{\prime}}\!(p^{\prime})
\Bigr],
\]
where the imaginary part of the retarded polarization tensor is \cite{blaizot2}
\small{
\[
{\rm Im}\Pi_{\mu\nu}^{(R)}(p)\!=\!
-\omega m_D^2\!\!
\int\!\!\frac{d\Omega_{{\bf v}}}{4\pi}
\!\!\int\!\!\frac{d\Omega_{{\bf v}^{\prime}}}{4\pi}
\!\!\int\!\!\frac{d\Omega_{{\bf v}_1}}{4\pi}
\!\!\int\!\!\frac{d\Omega_{{\bf v}_2}}{4\pi}\,
v_{\mu}v_{\nu}^{\prime}
\Bigl\langle v\Big|\,\frac{1}{v\cdot p +i\hat{\rm C}}
\,\Big|v_1\!\Bigr\rangle\,
{\cal C}({\bf v}_1,{\bf v}_2)
\Bigl\langle v^{\prime}\Big|\,\frac{1}{v\cdot p -i\hat{\rm C}}
\,\Big|v_2\!\Bigr\rangle.
\]
}\normalsize
Furthermore, if in propagators $\!\,^{\ast}\tilde{\cal
D}^{\nu\nu^{\prime}}(p^{\prime})$ and $\!\,^{\ast}\tilde{\cal
D}^{\lambda\lambda^{\prime}}\!(p^{\prime})$ we keep only
longitudinal part (\ref{eq:5t}), then expression (\ref{eq:5u}) is
written in the compact form (cp. with Eq.\,(7.11) in
\cite{markov1})
\begin{equation}
-\frac{(2\pi)^7}{\vert\check{\bf v}\vert}\,m_D^2
\Biggl(\frac{N_c\alpha_s}{2{\pi}^2}\Biggr)^{\!2}\!
\Biggl(\frac{C_2}{N_c}\Biggr)\!
\int\!\!dp\!\int\!\!dp^{\prime}
\,\biggl(\frac{\omega}{{\bf p}^2}\!\biggr)
\,{\rm Im}\Bigl(p^2\!\,^{\ast}{\!\Delta}^l(p)\Bigr)
\biggl(\frac{\omega^{\prime}}{{\bf p}^{{\prime}2}}\!\biggr)
\,{\rm Im}\Bigl(p^{{\prime}2}\!\,^{\ast}{\!\Delta}^{\ast l}(p^{\prime})\Bigr)
\label{eq:5i}
\end{equation}
\[
\times\,
\frac{1}{\omega^{{\prime}5}}\,
\Bigl|
\stackrel{\scriptscriptstyle{\,(1)}}{\displaystyle{K}}_{0\nu}
\!(\check{\bf v}\vert\,p,-p^{\prime})
\tilde{u}^{\nu}(p^{\prime})
\Bigr|^2
\delta(\check{v}\cdot(p-p^{\prime})).
\]
We note that the imaginary part of the polarization tensor in
Eq.\,(\ref{eq:5u}) appears in the combination
\[
\biggl(-\frac{2T}{\omega^{\prime}}\biggr)
\,{\rm Im}\Pi_{\nu^{\prime}\lambda^{\prime}}^{(R)}(p^{\prime}).
\]
This function is the simplest example of (2-point) function of a
new set of n-point amplitudes (additional to the
ultrasoft amplitudes $^{\ast}\Gamma_{\mu_1\ldots\mu_n}^{a_1\ldots a_n}$)
proposed by Guerin \cite{guerin2, guerin3} needed to fulfill the
Kubo-Martin-Schwinger constraints. The existence of these amplitudes is
related to the presence of the damping caused by the collision operator
$\hat{\rm C}$.

Unfortunately the third term on the right-hand side of Eq.\,(\ref{eq:3g})
generated by the fluctuation current $\zeta_{\mu}$ gives no possibility
to write out a complete expression for the energy loss
$(-dE^{(1)}\!/dx)_{\rm ul.soft}$ in a simpler
and clear form similar to equation (\ref{eq:5i}). In general case we
can proceed as follows. We expand the collision kernel (\ref{eq:2e}) in spherical
harmonics (normalized by unit)
\begin{equation}
{\cal C}({\bf v},{\bf v}^{\prime})=4\pi\gamma
\sum\limits_{l,m}\sum\limits_{l^{\prime},m^{\prime}}
c_{lm,\,l^{\prime}m^{\prime}}
Y_l^m({\bf v})
Y_{l^{\prime}}^{m^{\prime}}\!({\bf v}^{\prime}).
\label{eq:5o}
\end{equation}
By virtue of the fact that the spherical harmonics are eigenvectors of the
collision kernel, i.e.,
$\gamma^{-1}\hat{\rm C}Y_l^m({\bf v})=\lambda_l Y_l^m({\bf v})$, the
coefficients $c_{lm,\,l^{\prime}m^{\prime}}$ in the expansion (\ref{eq:5o}) are
simply expressed in terms of eigenvalues $\lambda_l$ of the collision operator
\begin{equation}
c_{lm,\,l^{\prime}m^{\prime}} =
(-1)^{l^{\prime}-m^{\prime}}\lambda_{l^{\prime}}\,
\delta_{m,-m^{\prime}}\,\delta_{l,\,l^{\prime}},
\label{eq:5p}
\end{equation}
where
\[
\lambda_l = 1 -\frac{2}{\pi}
\int\limits_{-1}^{+1}\!dz\,\frac{z^2}{\sqrt{1-z^2}}\,P_l(z)
\]
and $P_l(z)$ are Legendry polynomials.
The explicit expressions for $\lambda_l$ are \cite{arnold2}
\begin{equation}
\lambda_{2n}=1-2\Biggl[\frac{(2n)!}{2^{2n}(n!)^2}\Biggr]^2
\Biggl(1+\frac{1}{(2n)^2+2n-2}\Biggr)\geq 0,
\label{eq:5a}
\end{equation}
\[
\lambda_{2n+1}=1,\quad n=0,1,2, \ldots\,.
\hspace{3.89cm}
\]
Substituting the expansion (\ref{eq:5o}) in (\ref{eq:5y}), taking into
account (\ref{eq:5p}) and property $Y_l^{\ast m}({\bf v})=
(-1)^{l-m}Y_l^{-m}({\bf v})$ we can write the expression for the energy loss
(\ref{eq:5y})
in the form of an expansion in modules squared of ``partial amplitudes''
\[
\left(\!-\frac{dE^{(1)\,l}}{dx}\right)_{\!{\rm ul.soft}}\! =
-\frac{(2\pi)^6}{\vert\check{\bf v}\vert}\,T\gamma\,m_D^2
\Biggl(\frac{N_c\alpha_s}{2{\pi}^2}\Biggr)^{\!2}\!
\Biggl(\frac{C_2}{N_c}\Biggr)\!
\int\!\!dp\!\int\!\!dp^{\prime}
\,\biggl(\frac{\omega}{{\bf p}^2}\biggr)
\,{\rm Im}\Bigl(p^2\!\,^{\ast}{\!\Delta}^l(p)\Bigr)
\,\delta(\check{v}\cdot(p-p^{\prime}))
\]
\begin{equation}
\times\Biggl[\,
\sum\limits_{n=0}^{\infty}\sum\limits_{m=-(2n+1)}^{m=2n+1}
|R^{(m)}_{2n+1}(p,p^{\prime};\check{v})|^2 +
\sum\limits_{n=1}^{\infty}\lambda_{2n}\!\sum\limits_{m=-2n}^{m=2n}
|R^{(m)}_{2n}(p,p^{\prime};\check{v})|^2
\,\Biggr],
\label{eq:5s}
\end{equation}
where
\[
R^{(m)}_{l}(p,p^{\prime};\check{v}) =
\int\!\!d\Omega_{{\bf v}^{\prime}}{\cal R}_{0}(p,p^{\prime};\check{v},v^{\prime})
Y_l^m({\bf v}^{\prime})
\]
are coefficient functions in the expansion of ${\cal R}_0$ in spherical
harmonics with respect to unit vector ${\bf v}^{\prime}$. By virtue of
(\ref{eq:5a}) the
expression in square brackets in Eq.\,(\ref{eq:5s}) is positively definite and
therefore the right-hand side of (\ref{eq:5s}) has a positive sign under
condition ${\rm Im}\Bigl(p^2\!\,^{\ast}{\!\Delta}^l(p)\Bigr)\leq 0$.

Let us write out the expression for the energy
losses following from the current correction of the second order in $\nu$:
$J_{Q\mu}^{(2)a}[\nu](p)$ (Eqs.\,(\ref{eq:3j}), (\ref{eq:3k})). The reasoning
analogous above-mentioned leads to the following formula
\[
\left(\!-\frac{dE^{(2)\,l}}{dx}\right)_{\!{\rm ul.soft}}\! =
-\frac{2}{\vert\check{\bf v}\vert}\,\Bigl[(2\pi)^6T\gamma\,m_D^2\Bigr]^{\!2}
\Biggl(\frac{N_c\alpha_s}{2{\pi}^2}\Biggr)^{\!3}\!
\Biggl(\frac{C_2}{N_c}\Biggr)\!
\int\!\!dp\!\int\!\!dp^{\prime}\!\int\!\!dp^{\prime\prime}
\,\biggl(\frac{\omega}{{\bf p}^2}\biggr)
\,{\rm Im}\Bigl(p^2\!\,^{\ast}{\!\Delta}^l(p)\Bigr)
\]
\begin{equation}
\times\Biggl[\,
\sum\limits_{l,\,m}\sum\limits_{\tilde{l},\,\tilde{m}}
\lambda_l\lambda_{\tilde{l}}\,
\bigg\{
|R^{(m,\,\tilde{m})}_{l,\,\tilde{l}}
(p,p^{\prime},p^{\prime\prime};\check{v})|^2
+
|R^{(\tilde{m},\,m)}_{\tilde{l},\,l}
(p,p^{\prime\prime},p^{\prime};\check{v})|^2
\label{eq:5d}
\end{equation}
\[
+\,
{\rm Re}\,\Bigl(
R^{(m,\,\tilde{m})}_{l,\,\tilde{l}}
(p,p^{\prime},p^{\prime\prime};\check{v})
R^{\ast(\tilde{m},\,m)}_{\tilde{l},\,l}
(p,p^{\prime\prime},p^{\prime};\check{v})
\Bigr)\biggl\}\!\Biggl]
\,\delta(\check{v}\cdot(p-p^{\prime}-p^{\prime\prime})),
\]
where $\sum_{l,\,m}\equiv\sum_{l=1}^{\infty}\sum_{m=-l}^{m=l}$ and
\[
R^{(m,\tilde{m})}_{l,\,\tilde{l}}(p,p^{\prime},p^{\prime\prime};\check{v}) =
\int\!\!d\Omega_{{\bf v}^{\prime}}\!\int\!\!d\Omega_{{\bf v}^{\prime\prime}}
{\cal R}_{0}
(p,p^{\prime},p^{\prime\prime};\check{v},v^{\prime},v^{\prime\prime})
Y_l^m({\bf v}^{\prime})Y_{\tilde{l}}^{\ast\,\tilde{m}}({\bf v}^{\prime\prime}).
\]
In deriving  (\ref{eq:5d}) we consider that the function ${\cal R}_{\mu}
(p,p^{\prime},p^{\prime\prime};\check{v},v^{\prime},v^{\prime\prime})$
satisfies relation similar to (\ref{eq:4y}). Note that if we introduce a new
functions
\[
R^{(S)(m,\,\tilde{m})}_{l,\,\tilde{l}}
(p,p^{\prime},p^{\prime\prime};\check{v})
\equiv
\frac{1}{2}\,\Bigl(
R^{(m,\,\tilde{m})}_{l,\,\tilde{l}}
(p,p^{\prime},p^{\prime\prime};\check{v})
+
R^{(\tilde{m},\,m)}_{\tilde{l},\,l}
(p,p^{\prime\prime},p^{\prime};\check{v})
\Bigr),
\]
\[
R^{(A)(m,\,\tilde{m})}_{l,\,\tilde{l}}
(p,p^{\prime},p^{\prime\prime};\check{v})
\equiv
\frac{1}{2}\,\Bigl(
R^{(m,\,\tilde{m})}_{l,\,\tilde{l}}
(p,p^{\prime},p^{\prime\prime};\check{v})
-
R^{(\tilde{m},\,m)}_{\tilde{l},\,l}
(p,p^{\prime\prime},p^{\prime};\check{v})
\Bigr),
\]
then an expression within parentheses in integrand of Eq.\,(\ref{eq:5d})
is rewritten in the following form (for the sake of brevity we suppress
arguments of functions)
\[
|R^{(m,\,\tilde{m})}_{l,\,\tilde{l}}|^2
+
|R^{(\tilde{m},\,m)}_{\tilde{l},\,l}|^2
+
{\rm Re}\,\Bigl(
R^{(m,\,\tilde{m})}_{l,\,\tilde{l}}
R^{\ast(\tilde{m},\,m)}_{\tilde{l},\,l}
\Bigr)=
3|R^{(S)(m,\,\tilde{m})}_{l,\,\tilde{l}}|^2
+
|R^{(A)(m,\,\tilde{m})}_{l,\,\tilde{l}}|^2.
\]
The right-hand side of the last expression is explicitly positive
definite.

By using the explicit expressions for the energy losses, one can roughly
estimate their order in the coupling constant. For values $\omega\sim
g^4T,\,|{\bf p}|\sim g^2T$ from Eqs.\,(\ref{eq:5r}), (\ref{eq:5s}) and
(\ref{eq:5d}) the following estimations can be obtained:
\begin{equation}
\left(\!-\frac{dE^{(0)}}{dx}\right)_{\!{\rm ul.soft}}\!
\sim\alpha_s^4\,T^2,\quad
\left(\!-\frac{dE^{(1)}}{dx}\right)_{\!{\rm ul.soft}}\!
\sim\alpha_s^4\,T^2,\quad
\left(\!-\frac{dE^{(2)}}{dx}\right)_{\!{\rm ul.soft}}\!
\sim\alpha_s^4\,T^2
\label{eq:5f}
\end{equation}
up to the possible logarithmic factor $\ln(1/g)$. It is seen that at the
ultrasoft momentum scale the energy losses are parametrically
strongly suppressed for
$\alpha_s\!\ll\!1$ as compared with the losses induced by scattering of the
energetic parton off plasma excitations at the soft momentum scale
\cite{markov1}.
However the estimations (\ref{eq:5f}) suggest that not only the first three
terms, but all terms of higher orders will be values of the same order in the
coupling constant,
i.e., $(dE^{(n)}/dx)_{\rm ul.soft}\sim\alpha^4_s\,T^2,\,n>2$.
Thus from formal point of view for the estimation of real value of the
energy losses we must take into account the whole series
$\sum_{n=0}^{\infty}J_{Q \mu}^{(n)ab}[\nu](p)\,Q_0^b$ in initial expression
for the energy losses  (\ref{eq:5q}).

\section{\bf Conclusion}
\setcounter{equation}{0}

In this work the procedure of calculation of the effective current
generating the processes of the nonlinear interaction of the energetic
color particle with the ultrasoft gluon fluctuations was proposed. The
energy losses associated with these processes are also taken into
account. We have established that the contribution of each term in the
expansion of the effective current in powers of the noise term $\nu(X,{\bf
v})$ to the energy losses is strongly suppressed in the coupling\footnote{This
means that probably, the mechanism is irrelevant for phenomenology at RHIC.}
and all these contributions are of the same order in $g$. By virtue of
the last circumstance
perturbation approach for calculation of the energy losses breaks down
at the ultrasoft momentum scale. The complete calculation of the
energy losses $(dE\!/dx)_{\rm ul.soft}$ turns out to be non-perturbative
and thus it requires an all-order resummation that cannot be performed
via the methods developed here.

\section*{\bf Acknowledgments}
This work was supported by the Russian Foundation for Basic Research
(project no 03-02-16797).

\section*{Appendix}
\setcounter{equation}{0}

Here we present an explicit form of projection operators $P_{\mu\nu}(p)$
and $Q_{\mu \nu}(p)$ in two gauges:
\begin{itemize}
\item in the {\it covariant gauge}

$$
P_{\mu\nu}(p) = g_{\mu\nu}-D_{\mu\nu}(p)-Q_{\mu\nu}(p),\quad
Q_{\mu \nu}(p) =
\frac{\bar{u}_{\mu}(p)\bar{u}_{\nu}(p)}{\bar{u}^2(p)}\,,\quad
D_{\mu \nu}(p)=\frac{p_{\mu}p_{\nu}}{p^2}\,,\quad
$$
$$
\bar{u}_{\mu}=p^2u_{\mu}-p_{\mu}(p\cdot u),
$$
\item in the {\it temporal gauge}
$$
P_{\mu\nu}(p) = g_{\mu \nu}-u_{\mu}u_{\nu}-\frac{(p\cdot u)^2}{p^2}
\,\tilde{Q}_{\mu \nu}(p), \quad
\tilde{Q}_{\mu \nu}(p) =
\frac{\tilde{u}_{\mu}(p)\tilde{u}_{\nu}(p)}{\bar{u}^2(p)}\,,
$$
$$
\tilde{u}_{\mu}(p)=\frac{p^2}{(p\cdot u)}\,(p_{\mu}-u_{\mu}(p\cdot u)),
$$
\end{itemize}
where $u_{\mu}$ is global 4-velocity of the non-Abelian plasma. One assume
that we are in a rest frame of a heat bath, so that $u_{\mu}=(1,0,0,0).$

\end{document}